\documentclass[prd,twocolumn,superscriptaddress,altaffilletter,nofootinbib]{revtex4}

\usepackage{cancel}
\usepackage{graphicx}
\usepackage{amsmath}
\usepackage{amssymb}
\usepackage{graphicx,epsfig}
\newtheorem{theorem}{Theorem}

\newcommand{\be}{\begin{equation}}
\newcommand{\ee}{\end{equation}}
\newcommand{\bea}{\begin{eqnarray}}
\newcommand{\eea}{\end{eqnarray}}
\newcommand{\der}{\partial}
\newcommand{\vphi}{\varphi}

\begin{document}



\title{Gauge invariant wormholes}

\author{Israel Quiros}\email{iquiros@fisica.ugto.mx}\affiliation{Dpto. Ingenier\'ia Civil, Divisi\'on de Ingenier\'ia, Universidad de Guanajuato, Campus Guanajuato, Gto., M\'exico.}

\date{\today}

\begin{abstract} We aim at finding static, spherically symmetric, vacuum solutions of a gauge invariant theory of gravity over Weyl integrable geometry spaces. It arises that vacuum wormholes of pure geometric nature are solutions of this theory. This means that there is not necessary to place any exotic matter at the throat of the wormhole in order to support it. A thorough discussion of the related gauge freedom and its experimental consequences is also given.\end{abstract}


\maketitle




\section{Introduction}\label{sect-intro}


A promising geometrical arena where to investigate gauge symmetry in gravitational theories is Weyl geometry, i. e., in the dynamical formulation of gravitation theory one should replace Riemann by Weyl geometry. In the latter, vanishing of the covariant derivative of the metric is not required. Instead, the nonmetricity law driving the affine properties of Weyl space -- denoted here by $\tilde W_4$ -- reads:

\bea \hat\nabla_\alpha g_{\mu\nu}=-Q_\alpha g_{\mu\nu},\label{vect-nm}\eea where $Q_\mu$ is the nonmetricity vector and the covariant derivative $\hat\nabla_\mu$ is defined with respect to the torsion-free affine connection of the manifold: $\Gamma^\alpha_{\;\;\mu\nu}=\{^\alpha_{\mu\nu}\}+L^\alpha_{\;\;\mu\nu},$ where $\{^\alpha_{\mu\nu}\}:=g^{\alpha\lambda}\left(\der_\nu g_{\mu\lambda}+\der_\mu g_{\nu\lambda}-\der_\lambda g_{\mu\nu}\right)/2$, is the Levi-Civita (LC) connection, while $L^\alpha_{\;\;\mu\nu}:=\frac{1}{2}\left(Q_\mu\delta^\alpha_\nu+Q_\nu\delta^\alpha_\mu-Q^\alpha g_{\mu\nu}\right),$ is the disformation tensor. The Weyl gauge vector $Q_\mu$ measures how much the length of given vector varies during parallel transport. When the nonmetricity vector in \eqref{vect-nm} amounts to the gradient of a geometric scalar: $Q_\mu=\der_\mu\phi^2/\phi^2=2\der_\mu\phi/\phi,$ the nonmetricity law \eqref{vect-nm} is replaced by

\bea \hat\nabla_\alpha g_{\mu\nu}=-\frac{\der_\alpha\phi^2}{\phi^2}g_{\mu\nu}=-2\frac{\der_\alpha\phi}{\phi}g_{\mu\nu}.\label{grad-nm}\eea The spaces where the gradient nonmetricity condition \eqref{grad-nm} is satisfied are called as Weyl integrable geometry (WIG) spaces.\footnote{In this case the disformation tensor takes the form: $L^\alpha_{\;\;\mu\nu}:=\phi^{-1}\left(\der_\mu\phi\delta^\alpha_\nu+\der_\nu\phi\delta^\alpha_\mu-\der^\alpha\phi g_{\mu\nu}\right).$} Here we denote them by $\tilde W^\text{wig}_4$.

Local scale invariance is one of the most important properties of WIG spaces, due to invariance of the nonmetricity law \eqref{grad-nm}, and also of the affine connection and the related Weyl-Riemann curvature tensor (among other quantities,) under the Weyl gauge transformations:

\bea g_{\mu\nu}\rightarrow\Omega^2g_{\mu\nu},\;\phi\rightarrow\Omega^{-1}\phi,\;\Psi_A\rightarrow\Omega^{w_A}\Psi_A,\label{gauge-t}\eea where the $\Psi_A$ represent non gravitational fields with conformal weights $w_A$. In what follows, for simplicity, we shall call these just as ``gauge transformations.''


In recent papers \cite{quiros-prd-2023-1, quiros-arxiv} we have investigated, from a non-conventional perspective, the role gauge symmetry may play in the explanation of present and, perhaps, future gravitational phenomena. The statement of our research challenges the widespread belief that gauge symmetry, if it was a symmetry of the gravitational laws at all, is to be a pretty broken symmetry, so that it may have played a role at very early times in the cosmic evolution or in short range gravitational interactions \cite{smolin, cheng, bars, bars-sailing, ghilen-prd, ghilen-1, ghilen-2}. The belief is based on the break down of gauge symmetry by nonvanishing masses of particles. Hence, immediately after break down of electroweak (EW) $SU(2)\times U(1)$ --symmetry and consequent acquirement of masses by the particles of the standard model (SM), the assumed gauge symmetry of the gravitational laws should be broken as well. This means that gauge symmetry may have played a role up to the radiation-dominated stage of the cosmic evolution. 

The situation changed after publication of \cite{bars}, where it was shown that the SM of particles and fields can be modified in such a way as to be compatible with gauge symmetry. The secret was in lifting the mass parameter appearing in the symmetry breaking potential within the Higgs Lagrangian, to a field with appropriate transformation properties under the gauge transformations.  Means that the SM fields aquire point-dependent masses.

In \cite{quiros-prd-2023-1}, we have shown that the only gauge invariant theory of gravity which admits coupling of matter fields with nonvanishing masses, is given by the gravitational Lagrangian: ${\cal L}_\text{grav}=\phi^2\hat R/2-\lambda\phi^4/8$, where $\lambda$ is a free constant. This theory is based in WIG background spaces. Hence, in the above Lagrangian $\hat R$ is the curvature scalar of WIG space and $\phi$ is the Weyl gauge scalar (a scalar field of geometric nature.) Thanks to the decomposition of WIG geometric objects and operators in terms of their Riemannian equivalents we have that: $\hat R=R+6(\der\phi)^2/\phi^2-3(\nabla^2\phi^2)/\phi^2$, where we use the notation $(\der\phi)^2\equiv g^{\mu\nu}\der_\mu\phi\der_\nu\phi$ and $\nabla^2\equiv g^{\mu\nu}\nabla_\mu\nabla_\nu$. Hence, up to a total derivative the above gravitational Lagrangian can be written in the following way:

\bea {\cal L}_\text{grav}=\frac{1}{2}\left[\phi^2 R+6(\der\phi)^2-\frac{\lambda}{4}\phi^4\right],\label{grav-lag}\eea where $R$ is the standard Ricci scalar, i. e., it is the curvature scalar of Riemann space. Variation of the gravitational Lagrangian \eqref{grav-lag} with respect to the metric leads to the vacuum EOM:

\bea &&G_{\mu\nu}-\frac{1}{\phi^2}\left(\nabla_\mu\nabla_\nu-g_{\mu\nu}\nabla^2\right)\phi^2\nonumber\\
&&\;\;\;\;\;\;+\frac{6}{\phi^2}\left[\der_\mu\phi\der_\nu\phi-\frac{1}{2}g_{\mu\nu}(\der\phi)^2\right]=-\frac{\lambda}{8}\phi^2g_{\mu\nu},\label{vac-eom}\eea or if substitute its trace trace back into \eqref{vac-eom}, we obtain:

\bea &&{\cal R}_{\mu\nu}:=R_{\mu\nu}+\frac{6}{\phi^2}\der_\mu\phi\der_\nu\phi\nonumber\\
&&\;\;\;\;\;\;\;\;-\frac{1}{\phi^2}\left(\nabla_\mu\nabla_\nu+\frac{1}{2}g_{\mu\nu}\nabla^2\right)\phi^2=\frac{\lambda}{8}\phi^2g_{\mu\nu}.\label{vac-eq}\eea Meanwhile, variation of \eqref{grav-lag} with respect to the scalar field yields the ``Klein-Gordon'' EOM:

\bea R+6\frac{(\der\phi)^2}{\phi^2}-3\frac{\nabla^2\phi^2}{\phi^2}=\frac{\lambda}{2}\phi^2.\label{phi-eom}\eea This is not an independent equation since it coincides with the trace of \eqref{vac-eq}. This means that the scalar field EOM \eqref{phi-eom} is not an independent equation. In consequence $\phi$ does not obey an specific EOM: it is a free function which can be chosen at will. 

Given that $\phi$ is not a dynamical degree of freedom (DOF) the kinetic term for the scalar field in \eqref{grav-lag} does not affect the measured Newton's constant, as it does in Brans-Dicke (BD) type theories.\footnote{In the BD theory, since the scalar field $\phi$ is a dynamical DOF, the measured gravitational constant is modified not only by $\phi$, but also by the kinetic term through the BD coupling constant $\omega_\text{BD}$ \cite{bd-theory, fujii-book}: $$8\pi G_N=\frac{1}{\phi^2}\left(\frac{4+2\omega_\text{BD}}{3+2\omega_\text{BD}}\right).$$} The measured Newton's constant in the class of theories \eqref{grav-lag} corresponds to the tensor gravitational force. It is given by:

\bea 8\pi G_N(x)=M^{-2}_\text{pl}(x)=\frac{1}{\phi^2(x)},\label{newton-c}\eea where $M^2_\text{pl}(x)$ is the point-dependent squared effective Planck mass. For the same reason above: $\phi$ is not a dynamical DOF, the wrong sign of the kinetic energy density term for $\phi$ in ${\cal L}_\text{grav}$, is harmless.


The gauge invariant approach to gravity which is based in \eqref{grav-lag} over WIG spaces, complemented with the modification of the Higgs Lagrangian proposed in \cite{bars}, opens up the possibility for gauge symmetry to be an actual symmetry of the classical laws of gravity. The study of the observational cosmological consequences of this approach, was performed in \cite{quiros-arxiv}. In the latter bibliographic reference we have shown that gauge freedom -- the most immediate consequence of gauge symmetry -- is not innocuous and that, contrary to widespread belief, the gauge choice has observational cosmological consequences. 

The problem is that, if gauge choice has observational consequences, then the popular understanding of gauge fixing within gauge invariant gravitational theories is not adequate anymore. Actually, if observations prefer one gauge over the remaining infinity of possible gauges, this means that only one gauge yields the correct description of given gravitational phenomena. Hence, we can not base our choice of gauge neither on simplicity of mathematical handling nor on clarity of physical interpretation: Gauge fixing is not a trivial task anymore. Given the impact of this result it is required additional demonstration that the gauge choice can be submitted to experimental verification. In the present paper we shall show that local (terrestrial and Solar system) experiments are able to differentiate between the different gauges. This corroborates (and complements) previous results obtained within the cosmological context \cite{quiros-arxiv}. Means that, contrary to common wisdom, the gauge choice bears experimental and physical consequences. In Sections \ref{sect-gauges}, \ref{sect-vac-gauges}, \ref{sect-z} and \ref{discu-gauge-freedom}, we shall discuss the various aspects of this result in details.


The plan of the paper is as follows. In Section \ref{sect-gauges} special attention is paid to gauge freedom in connection with one of the main properties of this theory: gauge symmetry. In this Section we explain our non-conventional approach to gauge freedom. This approach has been formerly exposed in the bibliographic references \cite{quiros-prd-2023-1, quiros-arxiv}. In Section \ref{sect-theorem} we demonstrate a theorem about spherically symmetric solutions of the vacuum EOM \eqref{vac-eq}, which represents a generalization of the Birkhoff's theorem to include the class of gauge invariant theories given by \eqref{grav-lag}. In Section \ref{sect-sol} we find the only static, spherically symmetric solutions of the vacuum EOM \eqref{vac-eq}. Although these solutions are an immediate consequence of the above mentioned generalization of Birkhoff's theorem, in this section we derive the solutions based on the minimal set of additional assumptions, so that the derivation amounts to an additional proof of this theorem. Section \ref{sect-schw-ds} is dedicated to further generalizing the theorem, to the case of spherically symmetric de Sitter vacuum. By comparing the static, spherically symmetric line-element obtained for the vacuum case, with the generic wormhole line-element, in Section \ref{sect-wormholes} we show that there are possible wormhole solutions for the vacuum of the gauge invariant theory under study. This means that the resulting wormhole does not require of any exotic matter at the throat. In Section \ref{sect-vac-gauges} we explore three different gauges of the static, spherically symmetric solutions, which correspond to three different choices of the geometric scalar field. In section \ref{sect-z} we investigate the possible experimental consequences of the gauge choice. These are associated with measurements of the relative redshift of frequencies, which yield different results in the different gauges. We are able to make estimates of the magnitude of the redshift in the three gauges studied in Section \ref{sect-wormholes}. The results of this paper are discussed in Section \ref{sect-discu}, where we pay special attention to two main results: 1) the relevance of the generalization of Birkhoff's theorem proofed in Section \ref{sect-theorem}, and 2) the novelty of our approach to gauge freedom. The role of the gauge invariants in our approach is also discussed in Section \ref{sect-discu}. Brief conclusions are given in Section \ref{sect-conclu}.

In this paper we assume the mostly positive signature of the metric: $(-,+,+,+)$. Greek indexes ($\alpha$, $\beta$,$\cdots$, $\mu$, $\nu$, etc.) are spacetime indexes, so that these run from $0$ to $3$. Meanwhile lower case latin indexes $i,j,k,\cdots=1,2,3$, are spatial indexes.


\section{On gauge freedom}\label{sect-gauges}


The results of \cite{quiros-arxiv}, together with the results which we shall expose in Section \ref{sect-z} of the present paper, tell us that a different approach to gauge freedom is to be undertaken. In this section we shall explain the details of our non-conventional approach, which has been formerly explained in \cite{quiros-prd-2023-1, quiros-arxiv}.

The fact that, on the one hand, the geometric scalar field $\phi$ -- being a free function -- can be any smooth function $\phi=\phi_a(x)$, where $a=0,1,2,...,N$ ($N\rightarrow\infty$,) labels the different functions in the set of all real-valued functions while, on the other hand, $\phi$ determines the point-dependent Newton's constant which is measured in Cavendish-type experiments: $8\pi G_N=\phi^{-2}$, means that each choice of a function $\phi_a$ -- properly a gauge choice -- leads to a specific theory with its measured quantities. 

The general element of a gauge can be defined in the following way \cite{quiros-prd-2023-1}:

\bea {\cal G}_a=\{({\cal M}_4,g^{(a)}_{\mu\nu},\phi_a)|{\cal L}^{(a)}_\text{tot},{\cal S}_a,{\cal C},\cdots\},\label{gauge-a}\eea where ${\cal M}_4\in\tilde W_4^\text{wig}$ is a WIG spacetime manifold. In this expression ${\cal L}^{(a)}_\text{tot}={\cal L}^{(a)}_\text{grav}+{\cal L}^{(a)}_\text{sm}$, where ${\cal L}^{(a)}_\text{grav}$ is the gravitational Lagrangian \eqref{grav-lag} operating in the gauge ${\cal G}_a$:

\bea {\cal L}^{(a)}_\text{grav}=\frac{1}{2}\left[\phi_a^2 R_a+6(\der\phi_a)^2-\frac{\lambda}{4}\phi_a^4\right],\label{gauge-a-lag}\eea with $R_a\equiv R[g^{(a)}_{\mu\nu}]$, and ${\cal L}^{(a)}_\text{sm}$ is the Lagrangian of the standard model of particles and fields, including the modified gauge invariant Higgs Lagrangian \cite{bars}, specialized to ${\cal G}_a$. In equation \eqref{gauge-a-lag} ${\cal S}_a$ represents the set of point-dependent ``constants'' of Nature (for instance, the point-dependent Planck mass, $M^2_{\text{pl},a}=\phi_a^2$,) ${\cal C}$ is the set of gauge invariant constants (the Planck constant $\hbar$, the speed of light in vacuum $c$ and the quantum of electric charge $e$, among others,) while the ellipsis represent other relevant measured quantities of the theory as, for instance, the redshift. 

Due to gauge symmetry the collection of all possible gauges forms an equivalence class of conformally related theories, instead of a single theory.\footnote{Let us make a brief comparison with the well-known but hardly understood case of the conformal transformations in scalar-tensor theories (STT-s). In this case there is an infinity of possible representations of the given STT. Each representation is called a frame. The Jordan, Einstein and string frames, are among the most used ones in the bibliography. Any two frames are linked by a conformal transformation of the metric. One is tempted to identify the conformally related frames with the gauges in the conformal equivalence class ${\cal K}$. However, this would be incorrect, since gauge invariance is not a symmetry of the scalar-tensor theories of gravity. In consequence, there is not such a class of conformal equivalent theories. Notice that in the STT-s the scalar field satisfies a specific EOM so that it is a dynamical DOF.} The conformal equivalence class of theories can be expressed in the following way \cite{quiros-prd-2023-1}:

\bea {\cal K}=\left\{{\cal G}_0,{\cal G}_1,{\cal G}_2,...,{\cal G}_a,...,{\cal G}_N|\;a\in\mathbb{N}\right\},\label{ce-class}\eea where $N\rightarrow\infty$. Any two different gauges ${\cal G}_a$ and ${\cal G}_b$, in the conformal equivalence class ${\cal K}$, are related by the gauge transformation \eqref{gauge-t}:

\bea g^{(a)}_{\mu\nu}\rightarrow\Omega^2g^{(b)}_{\mu\nu},\;\phi_a\rightarrow\Omega^{-1}\phi_b,\label{gauge-t-ab}\eea or by the equivalent conformal transformation:

\bea g^{(a)}_{\mu\nu}\rightarrow\left(\frac{\phi_b}{\phi_a}\right)^2g^{(b)}_{\mu\nu}.\label{conf-t-ab}\eea One can move from one gauge to any other one and back by applying a gauge transformation \eqref{gauge-t-ab} and its inverse, respectively. Under the gauge transformation \eqref{gauge-t-ab} the gravitational Lagrangian ${\cal L}^{(a)}_\text{grav}$, transforms into 

\bea {\cal L}^{(b)}_\text{grav}=\frac{1}{2}\left[\phi_b^2 R_b+6(\der\phi_b)^2-\frac{\lambda}{4}\phi_b^4\right],\nonumber\eea while WIG space transforms into itself: $\tilde W^\text{wig}_4\rightarrow\tilde W^\text{wig}_4$. Means that the gravitational laws are invariant under \eqref{gauge-t-ab}. In other words: the laws of gravity look the same in any gauge, but for the GR gauge where these take the (simplest) GR form as shown below.


\subsection{General relativity gauge}\label{subsect-grg}


An outstanding gauge in the conformal equivalence class ${\cal K}$ defined in \eqref{ce-class}, which we shall identify as ${\cal G}_0$, is the one obtained if in \eqref{grav-lag} we make the following choice of the geometric scalar: $\phi(x)=\phi_0=$ const. Since it coincides with standard general relativity, we shall call this as ``general relativity gauge.'' The GR gauge can be obtained as well from any gauge ${\cal G}_a\in{\cal K}$ through the following gauge transformation \eqref{gauge-t-ab}:

\bea &&g^{(a)}_{\mu\nu}\rightarrow\Omega^2g^{(0)}_{\mu\nu},\;\phi_a\rightarrow\Omega^{-1}\phi_0,\nonumber\\
&&\Leftrightarrow\;g^{(a)}_{\mu\nu}\rightarrow\left(\frac{\phi_0}{\phi_a}\right)^2g^{(0)}_{\mu\nu},\label{gauge-t-a0}\eea where $g^{(0)}_{\mu\nu}=g^\text{gr}_{\mu\nu}$ is the metric which solves the GR equations of motion and $\phi_0=M_\text{pl}$ is the constant Plack mass (the measured Newton's constant $8\pi G_N=M_\text{pl}^{-2}$ is the one of GR theory.) Actually, under \eqref{gauge-t-a0}, the Lagrangian ${\cal L}^{(a)}_\text{grav}$ in \eqref{gauge-a-lag}, is transformed into the Einstein-Hilbert Lagrangian:\footnote{In the same way the SM Lagrangian ${\cal L}^{(a)}_\text{sm}$ is transformed into the standard Lagrangian ${\cal L}_\text{sm}$ which is not manifest gauge invariant.}

\bea {\cal L}_\text{eh}=\frac{M^2_\text{pl}}{2}\left(R_\text{gr}-2\Lambda\right),\label{eh-lag}\eea where $\Lambda=\lambda M^2_\text{pl}/8$ is the cosmological constant. The inverse transformation maps the Einstein-Hilbert Lagrangian ${\cal L}_\text{eh}$, as well as the SM Lagrangian ${\cal L}_\text{sm}$, back into the manifest gauge invariant Lagrangians ${\cal L}^{(a)}_\text{grav}$ and ${\cal L}^{(a)}_\text{sm}$. Notice that under \eqref{gauge-t-a0} Weyl integrable geometry space is transformed into Riemann space: $\tilde W^\text{wig}_4\rightarrow V_4$. Hence, the GR gauge ${\cal G}_0$ amounts to standard general relativity over Riemann background space. This way GR is just another gauge in the conformal equivalence class ${\cal K}$ defined in equation \eqref{ce-class}. Hence, even if GR itself is not manifest gauge invariant, it belongs in a larger class of gauge invariant gravitational theories.



\subsection{Additional comments}


The main difference of our approach with the most widespread understanding of gauge fixing is in the fact that, according to the latter, gauge fixing bears no physical consequences, since any gauge represents no more than a specific, complementary representation of the same theory. Meanwhile, according to our approach, any given gauge amounts to a different theory with its own set of measured quantities. 

Gauge invariance means that the laws of gravity \eqref{vac-eq} are the same in every gauge. Then, where the physical differences between the gauges lie in? The answer is: in the way which we perform measurements. Recall the metric tensor, the one which defines how to make measurements of time and distance, is different in the different gauges. The same is true for the density of fermions\footnote{In this regard notice that, for instance; the scalar density of fermions $\rho_\psi\propto\bar\psi\psi$, where $\psi$ is the fermion's spinor while $\bar\psi$ is its Dirac's adjoint, carries information about fermion's quantum state. It is transformed by the gauge transformations \eqref{gauge-t} in the following way: $\rho_\psi\rightarrow\Omega^{-3}\rho_\psi$, since the conformal weights of the fermion's spinor and of its Dirac's adjoint coincide: $w(\psi)=w(\bar\psi)=-3/2$.\label{fnote}} and also for the way the masses of the SM particles vary from point to point in spacetime. This is in addition to the already mentioned differences in the values (or functional forms) of the measured Newton's constant and in the results of the redshift measurements, as we shall show in Section \ref{sect-z}.


\subsection{Incorrect statements}


Let us make a comment about incorrect statements made in the bibliography in regard to the number of degrees of freedom and the choice of gauge in gravitational theories over WIG space. For instance, in the introductory part of Ref. \cite{ghilen-prd} (Section I.H), when analyzing the theory \eqref{grav-lag} with $\lambda=0$, it is stated that: i) the generation of the Planck scale as a vacuum expectation value (VEV) of $\phi$, by gauge fixing $\phi=M_\text{pl}$, demands the wrong sign kinetic energy term be present, and ii) when gauge fixing $\phi$ to a constant -- so that gauge symmetry is broken -- the number of DOF is not conserved. The latter is clearly a wrong statement since the scalar field $\phi$ is not a physical DOF. Recall that it does not obey an specific EOM. On the contrary, the gravitational DOF-s associated with the metric tensor: the two polarizations of the massless graviton, are physical degrees of freedom. Hence, before gauge fixing the number of DOF-s: $n_\text{dof}=2$, as in the GR gauge. Means that fixing the gauge, for instance: $\phi=M_\text{pl}$, does not change the number of physical DOF-s. For the same reason that $\phi$ is not associated with a physical DOF, the presence of the wrong sign kinetic energy term is harmless.


\section{A theorem}\label{sect-theorem}


In the next two sections we shall look for static, spherically symmetric solutions of the gauge invariant vacuum equations \eqref{vac-eq}. The fact that each such solution in one gauge generates a conformal solution in each one of the remaining gauges which belong in the same conformal equivalence class ${\cal K}$, simplifies the task. Let us, for simplicity, assume that in Eq. \eqref{vac-eq} the free constant $\lambda=0$. The following theorem takes place:

\begin{theorem} The only spherically symmetric solutions of the vacuum EOM-s \eqref{vac-eq} with $\lambda=0$: ${\cal R}_{\mu\nu}=0$, are conformal Schwarzschild solutions.\label{th-sols}\end{theorem} {\bf Proof.} As we have demonstrated general relativity is a specific gauge: ${\cal G}_0$, in the conformal equivalence class ${\cal K}$ defined in \eqref{ce-class}. Hence, any gauge ${\cal G}_a$ in ${\cal K}$ ($a=1,2,...,N$) is related with ${\cal G}_0$ through a gauge transformation \eqref{gauge-t-a0} and its inverse. Let us, for definiteness, set $\phi_0=1$ (this means that we work in units where $8\pi G_N=M^{-2}_\text{pl}=1$.) Hence, according to \eqref{gauge-t-a0}:

\bea g^{(a)}_{\mu\nu}=\phi_a^{-2}g^{(0)}_{\mu\nu}.\label{proof}\eea This transformation relates any gauge in ${\cal K}$ with the GR gauge and vice versa. Let us focus in the Schwarzschild metric in the GR gauge:

\bea g^\text{sch}_{\mu\nu}=\left[-\left(1-\frac{2m}{r}\right),\left(1-\frac{2m}{r}\right)^{-1},r^2,r^2\sin^2\theta\right],\label{sch-g}\eea which, according to the Birkhoff's theorem, is the only spherically symmetric solution of vacuum GR equations of motion. Hence, according to \eqref{proof} the only spherically symmetric solution of vacuum EOM derived from the gravitational Lagrangian \eqref{gauge-a-lag} (recall that we are considering $\lambda=0$,) representing a specific gauge ${\cal G}_a$, are given by:

\bea g^{(a),\text{vac}}_{\mu\nu}=\phi_a^{-2}g^\text{sch}_{\mu\nu}.\label{proof-1}\eea Since ${\cal G}_a$ can be any gauge in the conformal equivalence class ${\cal K}$, and since any element of the gauge invariant theory \eqref{grav-lag} must be in ${\cal K}$, then \eqref{proof-1} demonstrates the theorem \ref{th-sols}. {\bf Q.E.D.}

Theorem \ref{th-sols} is a generalization of Birkhoff's theorem to include the gauge invariant theory of gravity \eqref{grav-lag} with $\lambda=0$. In the next section we shall provide an additional proof of this theorem.


\section{Static, spherically symmetric solutions}\label{sect-sol}


Theorem \ref{th-sols} says that any spherically symmetric solution of the equations of motion \eqref{vac-eq} with $\lambda=0$, must be conformal to GR's Schwarzschild metric. In this section, temporarily, we shall forget about this theorem and we shall look for static, spherically symmetric solutions of vacuum EOM \eqref{vac-eq} with $\lambda=0$: ${\cal R}_{\mu\nu}=0$. Here we shall rely in the minimal set of assumptions, so that this will result into an independent proof of theorem \ref{th-sols}.

Let us write the most general static, spherically symmetric metric (we use spherical coordinates):

\bea ds^2=-e^{2\alpha(r)}dt^2+e^{2\beta(r)}dr^2+\rho^2(r)d\Omega^2,\label{local-met}\eea where $d\Omega^2\equiv d\theta^2+\sin^2\theta d\vphi^2$. The nonvanishing components of ${\cal R}_{\mu\nu}$ in \eqref{vac-eq} read (below we introduce the new variable $\chi\equiv\ln\phi$):

\bea &&{\cal R}_{00}=e^{2(\alpha-\beta)}\left[\alpha''+\alpha'^2-\alpha'\beta'+\frac{2\rho'}{\rho}\alpha'+\chi''\right.\nonumber\\
&&\left.\;\;\;\;\;\;\;\;\;\;\;\;+\left(3\alpha'-\beta'+\frac{2\rho'}{\rho}\right)\chi'+2\chi'^2\right]=0,\nonumber\\
&&{\cal R}_{rr}=-\alpha''-\alpha'^2+\alpha'\beta'+\frac{2\rho'}{\rho}\beta'-2\frac{\rho''}{\rho}-3\chi''\nonumber\\
&&\;\;\;\;\;\;\;\;\;\;\;\;-\left(\alpha'-3\beta'+\frac{2\rho'}{\rho}\right)\chi'=0,\nonumber\\
&&{\cal R}_{\theta\theta}=e^{-2\beta}\left[\rho\rho'\left(\beta'-\alpha'\right)-\rho'^2-\rho\rho''\right]+1\nonumber\\
&&\;\;-\rho^2e^{-2\beta}\left[\chi''+\left(\alpha'-\beta'+\frac{4\rho'}{\rho}\right)\chi'+2\chi'^2\right]=0,\nonumber\\
&&{\cal R}_{\vphi\vphi}={\cal R}_{\theta\theta}\sin^2\theta=0.\label{r-components}\eea The linear combination $e^{2(\beta-\alpha)}{\cal R}_{00}+{\cal R}_{rr}=0$, yields

\bea \chi''-\left(\alpha'+\beta'\right)\chi'-\chi'^2+\frac{\rho''}{\rho}-\frac{\rho'}{\rho}\left(\alpha'+\beta'\right)=0.\label{eq-1}\eea


\subsection{Family of conformal solutions}


Since there are four unknown functions: $\alpha$, $\beta$, $\rho$ and $\chi$, and only three differential equations in \eqref{r-components}, an additional condition on these unknowns may be introduced. Here we assume the following condition:

\bea \rho=e^{-\chi}r,\label{gauge}\eea which allows for great simplification of the third equation in \eqref{r-components}:

\bea {\cal R}_{\theta\theta}=e^{-2(\beta+\chi)}\left[r(\beta'-\alpha')-1\right]+1=0.\label{eq-2}\eea Besides, equation \eqref{eq-1} is greatly simplified as well:

\bea \alpha'+\beta'+2\chi'=0.\label{eq-1''}\eea As a consequence of condition \eqref{gauge}, the second equation in Eq. \eqref{r-components} is written in the following way:

\bea &&{\cal R}_{rr}=-\alpha''-\alpha'^2+\alpha'\beta'+\frac{2\beta'}{r}-\chi''\nonumber\\
&&\;\;\;\;\;\;\;\;\;\;\;+(\beta'-\alpha')\chi'+\frac{2\chi'}{r}=0.\label{eq-3}\eea Taking into account \eqref{eq-1''}, equation \eqref{eq-2} can be written as it follows:

\bea e^{-2(\beta+\chi)}\left[2r(\beta+\chi)'-1\right]+1=0.\label{eq-2'}\eea Substituting Eq. \eqref{eq-1''} into \eqref{eq-3} one gets:

\bea (\alpha+\chi)''+2(\alpha+\chi)'^2+2\frac{(\alpha+\chi)'}{r}=0.\label{eq-3'}\eea Equation \eqref{eq-2'} is equivalent to: $\left[re^{-2(\beta+\chi)}\right]'=1,$ which can be integrated to obtain:

\bea e^{2(\beta+\chi)}=\left(1+\frac{C_0}{r}\right)^{-1}.\label{sol-1}\eea where $C_0$ is an integration constant. Integration of equation \eqref{eq-3'} yields:

\bea e^{2(\alpha+\chi)}=C_2-\frac{C_1}{r},\label{sol-2}\eea where $C_1$ and $C_2$ are integration constants. Now, by integrating equation \eqref{eq-1''} it follows that: $\alpha+\beta+2\chi=B_0$, where $B_0$ is an integration constant. Hence, since multiplication of equations \eqref{sol-1} and \eqref{sol-2}:

\bea e^{2(\alpha+\beta+2\chi)}=C_2\left(1-\frac{C_1/C_2}{r}\right)\left(1+\frac{C_0}{r}\right)^{-1},\nonumber\eea it follows that $C_1=-C_0C_2$, with $C_2=e^{2B_0}$. The family of solutions is given by:

\bea &&e^{2\alpha}=C_2\,e^{-2\chi}\left(1+\frac{C_0}{r}\right),\nonumber\\
&&e^{2\beta}=e^{-2\chi}\left(1+\frac{C_0}{r}\right)^{-1},\label{sol-family}\eea or,

\bea &&ds^2=-e^{-2\chi}\left(1-\frac{2m}{r}\right)dt^2\nonumber\\
&&\;\;\;\;\;\;\;\;\;\;+e^{-2\chi}\left(1-\frac{2m}{r}\right)^{-1}dr^2+e^{-2\chi}r^2d\Omega^2,\label{line-e}\eea where we set the constant $B_0=0$ $\Rightarrow C_2=1$, and $C_0=-2m$. The line-element \eqref{line-e} can be rewritten in the following way: $ds^2=\phi^{-2}ds^2_\text{sch},$ where the Schwarzschild line-element reads; $ds^2_\text{sch}=g^\text{sch}_{\mu\nu}dx^\mu dx^\nu$, with the Schwarzschild metric given by \eqref{sch-g} and $x^\mu=(t,r,\theta,\vphi)$-- the spherical coordinates. This corroborates the theorem \ref{th-sols}.

The fact that the three differential equations in \eqref{r-components}, together with the additional condition \eqref{gauge}, were not enough to solve for the function $\chi=\ln\phi$, is a consequence of gauge invariance of the solution. Recall that the geometric scalar field $\phi$ does not obey a specific EOM due to invariance of \eqref{grav-lag} (and of the derived equations) under the gauge transformations \eqref{gauge-t}. The different gauges arise after fixing the scalar field $\phi$ to be any specific function of the coordinates (in the present case of the radial coordinate, exclusively): $\phi=\phi_a(r)$. Hence, in agreement with theorem \ref{th-sols}, the vacuum, spherically symmetric metric of any given gauge ${\cal G}_a$ is conformal to the Schwarzschild metric: $g^{(a)}_{\mu\nu}=\phi_a^{-2}g^\text{sch}_{\mu\nu}$.


\section{Generalization of theorem 1: de Sitter vacuum}\label{sect-schw-ds}


If in \eqref{vac-eq} we consider a nonvanishing $\lambda\neq 0$, then theorem \ref{th-sols} is to be modified. According to this theorem, which is a generalization of Birkhoff's theorem to include spherically symmetric solutions of gauge invariant (vacuum) theory of gravity \eqref{grav-lag}, \eqref{vac-eq} with vanishing $\lambda=0$: Any spherically symmetric solution of equations \eqref{vac-eq} with $\lambda=0$ must be, necessarily, conformal to Schwarzschild's solution.

In the present case, where we consider nonvanishing $\lambda$-term, the spherically symmetric GR solution is the known Schwarzschild-de Sitter metric. Hence, in theorem \ref{th-sols} one has to make the replacement ``Schwarzschild solution'' by ``Schwarzschild-de Sitter solution:'' 

\begin{theorem} The only spherically symmetric solutions of the de Sitter vacuum EOM-s \eqref{vac-eq}: ${\cal R}_{\mu\nu}=\lambda\phi^2 g_{\mu\nu}/8$, are conformal Schwarzschild-de Sitter solutions.\label{th-sols'}\end{theorem} {\bf Proof.} It is not difficult to proof that, in the case when $\lambda\neq 0$ in the vacuum equations \eqref{vac-eq}, the static, spherically symmetric solution is given by the following line-element:

\bea &&ds^2=e^{-2\chi}\left[-\left(1-\frac{2m}{r}-\frac{\Lambda}{3}r^2\right)dt^2\right.\nonumber\\
&&\left.\;\,\;\;\;\;+\left(1-\frac{2m}{r}-\frac{\Lambda}{3}r^2\right)^{-1}dr^2+r^2d\Omega^2\right],\label{schw-ds-line-e}\eea where $\Lambda=\lambda e^{B_0}/8$ and $B_0$ is an integration constant obtained after integrating equation \eqref{eq-1''}. In order to obtain this solution we relied on the fact that equation \eqref{eq-1} is still valid. Actually, for $\lambda\neq 0$ we have that:

\bea &&{\cal R}_{00}=-\frac{\lambda}{8}e^{2(\alpha+\chi)},\label{00-eom}\\
&&{\cal R}_{rr}=\frac{\lambda}{8}e^{2(\beta+\chi)}.\label{rr-eom}\eea Hence, the RHS of equation $e^{2(\beta-\alpha)}{\cal R}_{00}=-\lambda e^{2(\beta+\chi)}/8$, in the sum $e^{2(\beta-\alpha)}{\cal R}_{00}+{\cal R}_{rr}$, cancels out with the RHS of \eqref{rr-eom}. In consequence \eqref{eq-1} is to be satisfied so that, if take into account the condition \eqref{gauge}, then equation \eqref{eq-1''} is again obtained. By integrating out the latter equation it follows that: $2\chi=-\alpha-\beta+B_0$, where $B_0$ is the above mentioned integration constant. In what follows, without loss of generality, we set $B_0=0$. If in \eqref{00-eom} make the replacement: $\chi'\rightarrow-(\alpha'+\beta')/2$, and introduce the variable $u\equiv e^{\alpha-\beta}$, the following equation is obtained:

\bea u''+\frac{2}{r}u'=-\frac{\lambda}{4}.\label{u-eq}\eea The general solution of this equation reads: $u=1-2m/r-kr^2$, where $k=\lambda/24=\Lambda/3$. Then, since $\beta=-\alpha-2\chi$, we have that:

\bea &&u=e^{\alpha-\beta}=e^{2(\alpha+\chi)}=1-\frac{2m}{r}-\frac{\Lambda}{3}r^2\nonumber\\
&&\Rightarrow\;e^{2\alpha}=e^{-2\chi}\left(1-\frac{2m}{r}-\frac{\Lambda}{3}r^2\right).\nonumber\eea Finally, inserting the second line into the first one above we obtain:

\bea e^{2\beta}=e^{-2\chi}\left(1-\frac{2m}{r}-\frac{\Lambda}{3}r^2\right)^{-1}.\nonumber\eea {\bf Q.E.D.}

As it was in the case with $\lambda=0$, in the present case $\chi$ is a free function that is to be chosen at will. Different choices $\chi_a=\chi_a(r)$ $\Rightarrow\;\phi_a=\phi_a(r)$, define different gauges.


\section{Vacuum wormholes}\label{sect-wormholes} 


In this and in the next sections, for sake of simplicity, we shall consider the case with $\lambda=0$, exclusively. In equation \eqref{gauge} we have introduced a new variable which, when compared with \eqref{line-e}, happens to be the standard radial coordinate $\rho\equiv e^{-\chi}r$. In terms of this coordinate the line element \eqref{line-e} can be written in the following way:

\bea &&ds^2=-e^{-2\chi}\left(1-\frac{2m}{e^\chi\rho}\right)dt^2\nonumber\\
&&\;\;\;\;\;\;\;\;\;\;\;+\frac{\left(1+\rho\chi_{,\rho}\right)^2}{1-\frac{2m}{e^\chi\rho}}d\rho^2+\rho^2d\Omega^2,\label{line-e'}\eea where we used the notation $\chi_{,\rho}\equiv\der\chi/\der\rho$. In this form the line element reminds us the generic wormhole line element, which reads \cite{wh-morris, wh-lobo-2017};

\bea ds^2=-e^{2\Phi(\rho)}dt^2+\frac{d\rho^2}{1-\frac{b(\rho)}{\rho}}+\rho^2d\Omega^2,\label{wh-line-e}\eea where $\Phi(\rho)$ is the redshift function while $b(\rho)$ is the shape function \cite{wh-morris}. Hence, the class of geometries given by \eqref{line-e'} consists of vacuum wormholes of pure geometric nature since no matter is required in order to support them. This way the unwanted violation of energy conditions by matter at the throat of the wormhole is circumvented.

In order for the geometry depicted by \eqref{wh-line-e} to be a wormhole geometry, the functions $\Phi(\rho)$ and $b(\rho)$ have to fulfill several conditions (see below.) The interesting fact is that, in the present case, the wormhole metric is not fed by exotic matter at the throat but it is a solution of vacuum gravitational equations \eqref{r-components} in WIG background space $\tilde W_4^\text{wig}$. Consequently, the scalar field $\chi$ is not a matter field but it is a part of the geometry. As a matter of fact this field defines the nonmetricity of WIG space: $\hat\nabla_\alpha g_{\mu\nu}=-2\der_\alpha\chi g_{\mu\nu}$ (see equation \eqref{grad-nm}.)

Comparing equations \eqref{line-e'} and \eqref{wh-line-e} we find that the redshift and the shape functions are given by

\bea &&\Phi(\rho)=-\chi+\ln\sqrt{1-\frac{2m}{e^\chi\rho}},\nonumber\\
&&\frac{b(\rho)}{\rho}=\frac{2m+e^\chi\rho^2(2+\rho\chi_{,\rho})\chi_{,\rho}}{e^\chi\rho(1+\rho\chi_{,\rho})^2},\label{met-coef}\eea respectively. The proper radial distance is computed as,

\bea l(\rho)=\pm\int_{\rho_0}^\rho\frac{d\rho}{\sqrt{1-\frac{b(\rho)}{\rho}}}.\label{prop-dist}\eea It is required to be finite everywhere. The absence of horizons is warranted by fulfillment of the condition that $e^{2\Phi(r)}\neq 0$, so that also $\Phi(r)$ must be finite everywhere. 

The embedding diagram is useful to represent the wormhole and extract useful information to determine the shape function $b(\rho)$ \cite{wh-lobo-2017}. For this purpose we consider an equatorial slice $\theta=\pi/2$ and a fixed moment of time $t=$const. The line element \eqref{wh-line-e} simplifies to: $ds^2=d\rho^2/(1-b(\rho)/\rho)+\rho^2d\vphi^2$. In order to visualize the slice one embeds this metric into three-dimensional Euclidean space with metric: $ds^2=dz^2+d\rho^2+\rho^2d\vphi^2$, where $(\rho,\vphi,z)$ are the cylindrical coordinates. The embedded surface has equation $z=z(\rho)$, or in differential form \cite{wh-morris}: 

\bea \frac{dz}{d\rho}=\pm\frac{1}{\sqrt{\frac{\rho}{b(\rho)}-1}}.\label{embed-eq}\eea At the throat of the wormhole the function $\rho=\rho(r)$ is a minimum: $\rho=b(\rho)=\rho_\text{min}$. At this point the embedded surface is vertical: $dz/d\rho\rightarrow\infty$. On the contrary, far from the throat the spacetime is asymptotically flat: $dz/d\rho\rightarrow 0$. The flaring-out condition:

\bea \frac{d^2\rho}{dz^2}=\frac{b-\rho b_{,\rho}}{2b^2}>0,\label{flaring-out}\eea must be fulfilled at or near the throat at $\rho_\text{min}$. This is a fundamental ingredient of wormhole physics \cite{wh-lobo-2017}.

We want to mention that spherically symmetric vacuum BD wormholes \cite{wh-lobo-2010} differ from our vacuum wormhole solutions in that the corresponding line-element is not conformal to the Schwarzschild's solution. This does not entail violation of theorem \ref{th-sols} since, but for the singular value of the coupling constant $\omega_{BD}=-3/2$, BD theory is not a gauge invariant theory of gravity.


\section{Particular gauges of gravitational vacuum}\label{sect-vac-gauges}


In order to illustrate the gauge freedom property of theory \eqref{grav-lag} with $\lambda=0$, we shall make three different arbitrary choices of the function $\chi=\chi(r)$, among the infinity of possibilities. Recall, in this regard, that since the present theory is invariant under the gauge transformations \eqref{gauge-t}, the scalar field $\phi=e^\chi$ does not obey an specific EOM, i. e., it is not a physical DOF. Means that $\phi$ is a free function that may be chosen at will. Each choice defines a gauge.


\subsubsection{First gauge: GR theory}

  
The trivial choice is to set $\chi=0$ $\Rightarrow$ $\phi=1$. In this case we get the well-known spherically symmetric Schwarzschild black hole solution. We call the resulting description as the GR gauge ${\cal G}_0$ (see the related discussion in subsection \ref{subsect-grg},) which is the standard (Riemannian) geometric description resulting from considering vacuum GR with spherical symmetry.


\subsubsection{Second gauge: wormhole geometry}


Let us choose the gauge where;

\bea e^\chi=\sqrt{1-\frac{4m^2}{r^2}}=\sqrt{1-\frac{4m^2}{e^{2\chi}\rho^2}}.\label{gauge-2}\eea This equations can be written as: $x^2-x+4m^2/\rho^2=0$, where we set $x\equiv e^{2\chi}$. The roots of the above algebraic equation yield:

\bea e^{2\chi}=\frac{1}{2}\left(1\pm\sqrt{1-\frac{16m^2}{\rho^2}}\right).\label{e-chi}\eea In what follows we consider the positive branch exclusively, since the negative one leads to problems with positive definiteness of the redshift function. Since $\rho=e^{-\chi}r$, then

\bea \rho(r)=\frac{r^2}{\sqrt{r^2-4m^2}}\;\Rightarrow\;r=\frac{\sqrt{\rho^2+\rho\sqrt{\rho^2-16m^2}}}{\sqrt{2}}.\label{rho-r}\eea It follows that $\rho=\rho(r)$ is a minimum at $r=2\sqrt{2}m$, where $\rho_\text{min}=4m$. As $r$ decreases from infinity up to $r=2\sqrt{2}m$, the standard radial coordinate $\rho$ goes from infinity to its minimum value $\rho_\text{min}=4m$. As $r$ further decreases from $r=2\sqrt{2}m$ to $r=2m$, the radial coordinate $\rho$ grows from its minimum value $4m$ to infinity. The resulting spacetime geometry is free of singularity since the Kretschmann invariant: 

\bea K=\hat R^{\sigma\mu\nu\lambda}\hat R_{\sigma\mu\nu\lambda}=\frac{96m^2}{\rho^6\left(1+\sqrt{1-16m^2/\rho^2}\right)},\nonumber\eea where $\hat R^\sigma_{\;\;\mu\nu\lambda}$ is the curvature tensor of WIG space $\tilde W^\text{wig}_4$, is bounded: $0\leq K\leq 6/4^4m^4$.

For the redshift and shape functions: $\Phi=\Phi(\rho)$ and $b=b(\rho)$, we obtain the following expressions:

\bea &&e^{2\Phi}=\frac{2\sqrt\rho\left(\sqrt\rho\sqrt{\rho+\sqrt{\rho^2-16m^2}}-2\sqrt{2}m\right)}{\left(\rho+\sqrt{\rho^2-16m^2}\right)^{3/2}},\nonumber\\
&&\frac{b}{\rho}=\frac{2\sqrt{2}m\left(\rho^2-16m^2\right)\left(\rho+\sqrt{\rho^2-16m^2}\right)^{3/2}}{\sqrt\rho\left[\rho\left(\rho+\sqrt{\rho^2-16m^2}\right)-8m^2\right]^2}\nonumber\\
&&\;\;\;\;\;\;\;\;+\frac{8m^2\left[\left(\rho+\sqrt{\rho^2-16m^2}\right)^2-8m^2\right]}{\left[\rho\left(\rho+\sqrt{\rho^2-16m^2}\right)-8m^2\right]^2},\label{shape-g1}\eea respectively. It can be verified that the function $\Phi(\rho)$ is finite everywhere. In particular:

\bea \lim_{\rho\rightarrow\infty}e^{2\Phi}=1,\;\lim_{\rho\rightarrow 4m}e^{2\Phi}=2-\sqrt{2}.\nonumber\eea Besides, at the throat which is located at the minimum of the standard radial coordinate $\rho$: $\rho_\text{min}=4m$, we have that, $b(\rho)/\rho=1$ and $dz/d\rho\rightarrow\infty$, while as $\rho\rightarrow\infty$, $dz/d\rho\rightarrow 0$ (the space is asymptotically flat.) All of the remaining conditions required for the metric \eqref{wh-line-e} to represent a wormhole solution, including the flaring-out condition \eqref{flaring-out}, are satisfied as well.

In what follows we shall call this as the ``wormhole gauge.'' We want to underline that, depending of the specific choice of the function $\phi=\phi(r)$, there can be an infinity of possible wormhole configurations, i. e., of different wormhole gauges.


\subsubsection{Third gauge: naked singularity}


The third gauge corresponds to the following choice:

\bea e^\chi=\sqrt\frac{r}{r-2m}.\label{gauge-3}\eea In this case:

\bea \rho(r)=\sqrt{r(r-2m)},\nonumber\eea whose domain is $2m\leq r<\infty$ and its range: $0\leq\rho<\infty$. The line-element can be written in the following way:

\bea ds^2=-\frac{\rho^4dt^2}{\left(m+\sqrt{m^2+\rho^2}\right)^4}+\frac{\rho^2d\rho^2}{m^2+\rho^2}+\rho^2d\Omega^2.\label{line-e-nsing}\eea It is easily checked that this is not a wormhole geometry given that: 1) $\rho$ is not an extremum is its domain, 2) there is no throat and 3) the flaring-out condition is not fulfilled since: $d^2\rho/dz^2=-\rho/m^2<0$, among others. The metric \eqref{line-e-nsing} has a naked singularity at the origin $\rho=0$, where the Kretschmann invariant: 

\bea K=\frac{48m^2}{\rho^4\left(m+\sqrt{m^2+\rho^2}\right)^2},\nonumber\eea blows up. Below we shall call this as the ``naked singularity gauge.''

As seen, despite that the laws of gravity \eqref{vac-eq} (with $\lambda=0$) look the same in any gauge, each one of the chosen gauges represents a different physical/geometrical description of vacuum gravity. This happens because of the different ways of making measurements of time intervals, distances and massess, among others, in the different gauges. In the next section we shall demonstrate that these differences bear, in turn, observational differences. This means that experiment is able to pick out a gauge among the infinity of them: the one which better describes the amount of observational evidence.


\section{Redshift effect}\label{sect-z}


The distinctive property of WIG geometry space $\tilde W^\text{wig}_4$, is that the length of vectors varies from point to point in spacetime \cite{weyl-1918, many-weyl-book, london-1927, dirac-1973, utiyama-1973, adler-book, maeder-1978}. Yet, given that the units of measure change from point to point in the same way as the measured quantity does, by means of local measurements there is no way in which the measurements performed in one gauge can be differentiated from measurements in a different gauge. But, what about measurements which imply propagation of photons from one spacetime point to a distant one? Let us analyze, in particular, the gravitational shift of frequency. In this case a photon with certain frequency is emitted at some point with a given value of the gravitational potential and it is then detected at some other point with a different value of the gravitational potential. 

As we shall see, there are two different sources of frequency shift in WIG space: 1) the standard curvature shift which is due to the influence of the curvature of background space on the propagation of photons (this may be computed with the help of the null-geodesic equations,) and 2) the shift of frequency due to the variation of atomic transition energies in spacetime, which we call as ``nonmetricity'' shift of frequency. While in the GR gauge, which is associated with Riemann background space $V_4$, only the first kind of frequency shift arises, in any other gauge both types of shift occur.


\subsection{Curvature redshift}


The standard gravitational shift of frequency is described by the null-geodesic's equation:

\bea \frac{dk^\mu}{d\xi}+\left\{^\mu_{\nu\sigma}\right\}k^\mu k^\sigma=0,\label{0-geod}\eea where $k^\mu\equiv dx^\mu/d\xi$ is the wave-vector: $k_\mu k^\mu=0$, and $\xi$ is an affine parameter along null-geodesic. Based on dimensional analysis it can be checked that $k^\mu$ has conformal weight $w=-2$, like the fourth-momentum $p^\mu=mdx^\mu/ds$. As a consequence, the null-geodesic equation is not transformed by the gauge transformations \eqref{gauge-t} (see Appendix D of \cite{wald-book}.) This means that, such as it is with the gravitational laws \eqref{vac-eq}, the photon's Riemannian EOM \eqref{0-geod} is gauge invariant as well. In other words, photons (and radiation in general) follow null-geodesics of Riemann space $V_4$. Another way of saying this is that photons are ``blind'' to the structure of WIG space $\tilde W^\text{wig}_4$. 

Let us to start by describing the gravitational redshift in the GR gauge. Let us assume, for definiteness, that in the GR gauge ${\cal G}_0$ the photon propagates in the radial direction in a static, spherically symmetric Schwarzschild spacetime with metric \eqref{sch-g}: 

\bea ds_{(0)}^2=-A^2dt^2+A^{-2}dr^2+r^2d\Omega^2,\label{0-line-e}\eea where $A^2=-g^{(0)}_{00}=1-2m/r$. Let us further assume that the photon follows the radial direction with angular coordinates $(\theta,\vphi)=(\pi/2,0)$. Then, the components of the wave-vector of the photon in ${\cal G}_0$, read: 

\bea k_{(0)}^\mu\equiv\frac{dx^\mu}{d\xi_{(0)}}\;\Rightarrow\;k_{(0)}^\mu=\left(\omega_{(0)}, k_{r(0)},0,0\right),\label{wave-v-comp}\eea where $\omega_{(0)}=dt/d\xi_{(0)}$ and $k_{r(0)}=dr/d\xi_{(0)}$. Besides $k^{(0)}_\mu k_{(0)}^\mu=-A^2\omega_{(0)}^2+A^{-2}k^2_{r(0)}=0$. The null geodesics \eqref{0-geod} reduce to the following equations: 

\bea \frac{d\omega_{(0)}}{\omega_{(0)}}=-2\frac{dA}{A},\;\frac{dk_{r(0)}}{d\xi_{(0)}}=0,\label{0-geod'}\eea where we have taken into account that $k_{r(0)}dA/dr=dA/d\xi_{(0)}$. Straightforward integration of the first equation above leads to: $\omega_{(0)}=\omega_0/A^2=\omega_0/(-g^{(0)}_{00})$, where $\omega_0$ is an integration constant. The physical cyclic frequency reads: $\omega^{(0)}_\text{ph}=\sqrt{-g^{(0)}_{00}}\omega_{(0)}$, hence:

\bea \omega^{(0)}_\text{ph}(r)=\frac{\omega_0}{\sqrt{-g^{(0)}_{00}(r)}}=\frac{\omega_0}{\sqrt{1-2m/r}}.\label{phys-w-0}\eea 

Suppose that the photon is emitted at some time $t$ by an hydrogen atom placed at spatial point $(r,\pi/2,0)$ and is then absorbed at a later time $t_0$ by an identical hydrogen atom placed at $(r_0,\pi/2,0)$. Due to the effect of curvature on the propagation of the photon, there will be a relative shift of photon's frequency: 

\bea z^{(0)}_\text{curv}=\frac{\nu_{(0)}^\text{em}-\nu_{(0)}^\text{abs}}{\nu_{(0)}^\text{abs}}=\frac{\nu_{(0)}^\text{em}}{\nu_{(0)}^\text{abs}}-1,\label{z-curv-gen}\eea where $\nu_{(0)}^\text{em}\equiv\omega^{(0)}_\text{ph}(r)/2\pi$ is the measured frequency of the emitted photon, $\nu_{(0)}^\text{abs}\equiv\omega^{(0)}_\text{ph}(r_0)/2\pi$ is its frequency when it is absorbed by the second atom and $z^{(0)}_\text{curv}$ is the relative (curvature) ``redshift'' of frequency in the GR gauge. Hence, the photon is absorbed by the second hydrogen atom only if it is placed in a centrifuge with controlled rotation speed, as in Mossbauer experiment. In the GR gauge it is the only source of shift of photon's frequency. Then, according to \eqref{phys-w-0} and to \eqref{z-curv-gen}, the redshift of frequency in ${\cal G}_0$ is given by:

\bea z^{(0)}_\text{curv}=\frac{\sqrt{-g^{(0)}_{00}(r_0)}}{\sqrt{-g^{(0)}_{00}(r)}}-1.\label{z-curv-0}\eea Since the Riemannian null-geodesic equations \eqref{0-geod} are not modified by the gauge transformations \eqref{gauge-t}, then the same curvature shift of frequency arises in any other gauge ${\cal G}_a$:

\bea z^{(a)}_\text{curv}=\frac{\sqrt{-g^{(a)}_{00}(r_0)}}{\sqrt{-g^{(a)}_{00}(r)}}-1.\label{z-curv-a}\eea The following relationship between the curvature redshift in the GR gauge ${\cal G}_0$ and the same measured quantity in any other gauge ${\cal G}_a$, arises:

\bea z_\text{curv}=\frac{\phi(r)}{\phi_0}z_\text{gr}+\frac{\phi(r)}{\phi_0}-1,\label{z-rel}\eea where we took into account the theorem \ref{th-sols} expressed through equation \eqref{proof} and $\phi_0=\phi(r_0)$. Besides, in \eqref{z-rel} we dropped the index ``$a$,'' which denotes a specific gauge and we also renamed the curvature redshift taking place in the GR theory: $z_\text{gr}\equiv z^{(0)}_\text{curv}$.


\subsection{Nonmetricity redshift}


Unlike photons and radiation, time-like particles follow geodesics or autoparalles of WIG space $\tilde W^\text{wig}_4$. The autoparallel curves satisfy (for a detailed exposition see \cite{quiros-arxiv, quiros-prd-2022}):

\bea \frac{d^2x^\alpha}{ds^2}+\left\{^\alpha_{\mu\nu}\right\}\frac{dx^\mu}{ds}\frac{dx^\nu}{ds}-\frac{\der_\mu \phi}{\phi}h^{\mu\alpha}=0,\label{time-auto-p}\eea where

\bea h^{\mu\alpha}:=g^{\mu\alpha}+u^\mu u^\alpha=g^{\mu\alpha}-\frac{dx^\mu}{ds}\frac{dx^\alpha}{ds},\label{orto-proj}\eea is the orthogonal projection tensor, which projects any vector or tensor onto the hypersurface orthogonal to the four-velocity vector $u^\mu=dx^\mu/d\tau$. 

In $\tilde W^\text{wig}_4$ space, since the mass is a point-dependent quantity, then $m$ can not be taken out of the action integral: $S=\int mds.$ From this action the following EOM can be derived:

\bea \frac{d^2x^\alpha}{ds^2}+\left\{^\alpha_{\mu\nu}\right\}\frac{dx^\mu}{ds}\frac{dx^\nu}{ds}-\frac{\der_\mu m}{m}h^{\mu\alpha}=0,\label{time-geod}\eea where the non-Riemannian term $\propto\der_\mu m/m$ accounts for the variation of mass during parallel transport. Hence, since time-like autoparallels \eqref{time-auto-p} and time-like geodesics \eqref{time-geod} coincide in $\tilde W^\text{wig}_4$ space, we get that: 

\bea \frac{dm}{m}=\der_\mu\ln\phi dx^\mu\;\Rightarrow\;\frac{\der_\mu m}{m}=\frac{\der_\mu\phi}{\phi}.\label{mass-autop}\eea This equation can be readily integrated to get:

\bea m(r)=m_0\frac{\phi(r)}{\phi_0},\label{mass-phi-rel}\eea where $m_0=m(r_0)$ and $\phi_0=\phi(r_0)$ are the magnitudes of the mass and of the geometric scalar field evaluated at the value $r_0$ of the radial coordinate, respectively. Equation \eqref{mass-phi-rel} expresses how the mass of given time-like particle varies form point to point in $\tilde W^\text{wig}_4$ spacetime. This would lead, in particular, to a change in the atomic transition energies in spacetime. 

According to \eqref{mass-phi-rel} the masses of particles, for instance of the electron: $m_e$, vary from point to point in spacetime: $m_e(r)=m_{e0}\phi(r)/\phi_0$, where $m_{e0}$ is the value of the electron's mass at some reference point with spatial coordinates: $x^i_0=(r_0,\pi/2,0)$. Let $\nu_{if}$ be the frequency of a photon emitted by an hydrogen atom located at some spatial point with coordinates $x^i=(r,\pi/2,0)$, due to a transition from a state with principal quantum number $n_i$ into a state with $n_f$:

\bea \nu_{if}(r)=\frac{m_e(r)\alpha^2}{2}\left|\frac{1}{n_f^2}-\frac{1}{n_i^2}\right|,\label{nu-x}\eea where $\alpha$ is the fine structure constant. The frequency of the similar photon emitted/absorbed by an hydrogen atom placed at the reference point $x^i_0$ is given by: $\nu_{if}(r_0)=m_{e0}\alpha^2|n_f^{-2}-n_i^{-2}|/2$. Further assume that a photon emitted by an hydrogen atom placed at $(r,\pi/2,0)$, with frequency $\nu^\text{em}_{if}=\nu_{if}(r)$, is then absorbed by an hydrogen atom placed at $(r_0,\pi/2,0)$. No matter whether the photon's frequency is modified or not during its propagation, there is a redshift of frequency $z_\text{nm}$ which is associated, exclusively, with variation of the atomic transition energies in spacetime:

\bea z_\text{nm}=\frac{\nu^\text{em}_{if}}{\nu^\text{abs}_{if}}-1=\frac{m_e(r)}{m_{e0}}-1=\frac{\phi(r)}{\phi_0}-1,\label{z-nm}\eea where $\nu^\text{abs}_{if}=\nu_{if}(r_0)$. This shift of frequency is due to the nonmetricity of WIG space $\tilde W^\text{wig}_4$. As seen from \eqref{z-nm}, it does not depend on the specific atom which emits or absorbs the photon. Notice that in Riemann space $V_4$, since $\phi=\phi_0$ is a constant over all of spacetime, then: $z_\text{nm}=0$. 

In a WIG spacetime the overall redshift is a sum of the curvature and of the nonmetricity redshifts: $z_\text{tot}=z_\text{curv}+z_\text{nm}$, or if take into account \eqref{z-rel}, we can write the overall redshift of frequency in terms of the curvature redshift $z_\text{gr}$ taking place in the GR gauge:

\bea z_\text{tot}=\frac{\phi(r)}{\phi_0}z_\text{gr}+2\left[\frac{\phi(r)}{\phi_0}-1\right].\label{z-tot}\eea We recall that this equation is independent of the atom which emits/absorbs the photon.


\subsection{Experimental differentiation of gauges}


Equation \eqref{z-tot} relates the measured redshift in the GR gauge with the measured redshift in another gauge in the conformal equivalence class ${\cal K}$ of the theory \eqref{grav-lag} over background WIG space. This means that we can experimentally differentiate the different gauges through redshift experiments \cite{pound-rebka, pound-snider, vessot, will-lrr}.

For definiteness and in order to make the discussion more transparent, let us assume that measurement of the redshift is performed in a ``Pound-Rebka-type'' experiment \cite{pound-rebka, pound-snider}. A target $T_h$ made of some atom is placed at some height $h$ on top of a tower over the Earth surface (the radial coordinate $r=R_\oplus+h$, where $R_\oplus$ is the mean radius of the Earth) and an identical target $T_0$ (made of same atom) is placed at the bottom of the tower (radial coordinate $r_0=R_\oplus$.) Atoms in $T_h$ are in some excited state, so hat these are able to emit photons to return to the ground state. Emitted photons are then absorbed by atoms in the target $T_0$ which is located at the bottom of the tower.\footnote{Here we omit experimental details such as, for instance, that the target at the bottom of the tower must be mounted in a centrifuge with controlled rotation speed, etc.}


\subsubsection{General relativity gauge}


In the linear approximation, the gravitational redshift measured in the GR gauge ${\cal G}_0$ is given by \eqref{z-curv-0}:

\bea z_\text{gr}=\frac{\sqrt{1-\frac{2m}{R_\oplus}}}{\sqrt{1-\frac{2m}{R_\oplus+h}}}-1\approx-\frac{gh}{c^2},\label{z-gr}\eea where we took into account that $m\equiv G_NM_\oplus/c^2$ ($M_\oplus$ is the mass of the Earth) and that $g\equiv G_NM_\oplus/R^2_\oplus$. Hence, if choose the following experimental values: $h=1.00\times 10^2\,m$, $R_\oplus=6.371\times 10^6\,m$, $g=9.807\,m/s^2$ and $c=2.998\times 10^8\,m/s$, the magnitude of the gravitational redshift in general relativity is: $|z_\text{gr}|\approx 1.091\times 10^{-14}$.


\subsubsection{Wormhole gauge}


Let us compute the quantity $\phi(r)/\phi_0$ for the second gauge above: the wormhole gauge (here we assume the quadratic approximation since this would be the order of the modification of the GR redshift.) According to \eqref{gauge-2} we have that:

\bea \frac{\phi(r)}{\phi(r_0)}=\frac{\sqrt{1-\frac{4m^2}{(R_\oplus+h)^2}}}{\sqrt{1-\frac{4m^2}{R^2_\oplus}}}\approx 1+4z_\text{gr}^2\left(\frac{R_\oplus}{h}\right),\label{z-phi-g2}\eea where $z_\text{gr}$ is given by \eqref{z-gr}. If we substitute this equation back into \eqref{z-tot} in order to get the magnitude of the redshift measured in the wormhole gauge, we obtain:

\bea z_\text{tot}\approx z_\text{gr}+8z_\text{gr}^2\left(\frac{R_\oplus}{h}\right)=(1+\alpha)z_\text{gr},\label{z-tot-g2}\eea where we set: $\alpha=8z_\text{gr}(R_\oplus/h)$. For the chosen value of the height of the experimental tower ($h=100\,m$,) we get that $|\alpha|=5.560\times 10^{-9}$. This is far below the limit $|\alpha|<2\times 10^{-4}$, which was obtained in the most precise standard redshift test to date \cite{vessot, will-lrr}. This means that with the current precision of redshift experiments (this includes the experiments on local position invariance \cite{will-lrr},) it is not possible to differentiate the GR and the wormhole gauges.


\subsubsection{Naked singularity gauge}


In this case the gauge selection if defined by \eqref{gauge-3}:

\bea \frac{\phi(r)}{\phi(r_0)}=\sqrt\frac{1-\frac{2m}{R_\oplus}}{1-\frac{2m}{R_\oplus+h}}\approx 1+z_\text{gr},\label{z-phi-g3}\eea where, according to \eqref{z-gr}: $z_\text{gr}=-gh/c^2$, is the gravitational redshift in the GR gauge. The overall redshift in the ``naked singularity gauge'' is given by \eqref{z-tot}, which in the linear approximation reads:

\bea z_\text{tot}=(1+z_\text{gr})z_\text{gr}+2z_\text{gr}\approx 3z_\text{gr}.\label{z-tot-g3}\eea Means that this gauge is ruled out by redshift experiments \cite{will-lrr}.


\section{Discussion}\label{sect-discu}


We shall focus our discussion into two main results: i) the relevance of theorems \ref{th-sols}, \ref{th-sols'} and ii) the novelty of our interpretation of gauge freedom. The former is understood as a generalization of Birkhoff's theorem applied to gauge invariant theories of the class \eqref{grav-lag}, while the latter gains importance on the light of the widespread belief that the different gauges are dynamically and geometrically equivalent, so that these can not be experimentally differentiated. For completeness of this paper, at the end of section, we include a detailed discussion about the role of the gauge invariant quantities in the present approach.


\subsection{Relevance of the theorems \ref{th-sols} and \ref{th-sols'}}


Theorems \ref{th-sols} and \ref{th-sols'} are as useful in the search for spherically symmetric vacuum solutions of gauge invariant theory of gravity \eqref{grav-lag}, as it is the Birkhoff's theorem for spherically symmetric solutions of vacuum general relativity: If $\lambda=0$, any such spherically symmetric solution of equations \eqref{vac-eq} must be, necessarily, conformal to Schwarzschild's solution. In case $\lambda\neq 0$, any spherically symmetric solution of vacuum equations \eqref{vac-eq}, must be necessarily conformal to the Schwarzschild-de Sitter metric of the GR gauge.

These theorems rule out large classes of solutions. For instance, in the work of Ref. \cite{harko-prd-2023}, the authors investigated the static, spherically symmetric vacuum solutions of a gauge invariant theory developed in \cite{ghilen-1, ghilen-2}. After a specific choice of the nonmetricity vector: $Q_\mu=(0,Q_r,0,0)$ in \cite{harko-prd-2023}, the equations of motion of the mentioned theory result in the vacuum EOM \eqref{vac-eq} with nonvanishing $\lambda$-term. This means that theorem \ref{th-sols'} is valid in this case. The solution found by the authors of that reference (see also \cite{harko-epjc-2022}):

\bea &&e^{2\alpha}=e^{-2\beta}=1-\delta+\frac{\delta(2-\delta)}{6m}r-\frac{2m}{r}+C_3r^2,\nonumber\\
&&\phi(r)=\frac{\sqrt{C_1}}{C_2+r},\label{harko-sol}\eea where $C_1$, $C_2$ and $C_3$ are integration constants, does not satisfy theorem \ref{th-sols'} so that the solution \eqref{harko-sol} must be wrong. 

The main idea behind theorems \ref{th-sols} and \ref{th-sols'}, in this case, may be explained in the following way. As discussed in Section \ref{sect-gauges}, under the conformal transformation $g_{\mu\nu}=\phi^{-2}g^{(0)}_{\mu\nu}$, the gravitational Lagrangian \eqref{grav-lag} is transformed into the Einstein-Hilbert GR Lagrangian \eqref{eh-lag}. Hence, if the solutions \eqref{harko-sol} were correct, the following metric:

\bea ds^2=\phi^2(r)\left(-e^{2\alpha}dt^2+e^{2\beta}dr^2+r^2d\Omega^2\right),\nonumber\eea with $e^{2\alpha}$, $e^{2\beta}$ and $\phi$ given by \eqref{harko-sol}, were also a static, spherically symmetric solution of Einstein-de Sitter vacuum equations. But this is forbidden by Birkhoff's theorem (take the limit $\Lambda\rightarrow 0$.) A detailed discussion of the wrong assumptions and results of \cite{harko-prd-2023} is given in \cite{quiros-comment-2023}.


\subsection{Understanding gravitational gauge freedom}\label{discu-gauge-freedom}


One of the most controversial aspects of the present paper, which is shared with \cite{quiros-prd-2023-1, quiros-arxiv}, has to do with our novel interpretation of gauge freedom in gravitational theories. Usually gauge freedom is associated with dynamical and full physical equivalence of the different gauges. Each gauge offers different but complementary descriptions of given phenomenon. It seems that choosing a specific gauge is a matter of mathematical simplicity or of a clearer physical interpretation of the results. Means that the choice of gauge bears no physical and/or experimental consequences. Quite the opposite, here we have demonstrated that gauge choice is not innocuous: it has physical and experimental consequences. Actually, as we have discussed in Section \ref{sect-z}, local experiments can rule out given gauges. 

The widespread interpretation of gauge freedom as having neither physical nor experimental consequences, may be due to unconscious extrapolation of our understanding of gauge symmetry in field theory. Take, for instance, the electromagnetic (EM) $U(1)$ gauge symmetry. There are clear differences between gauge invariance within a gravitational theory and $U(1)$ gauge invariance. In the latter case Maxwell's and Dirac's equations are invariant under the $U(1)$ transformations:

\bea &&\psi\rightarrow e^{-ie\lambda(x)}\psi,\;\bar\psi\rightarrow e^{ie\lambda(x)}\bar\psi,\nonumber\\
&&A_\mu\rightarrow A_\mu+\der_\mu\lambda(x),\label{u1-t}\eea where $A_\mu$ is the EM vector potential, $\psi$ is the fermion's spinor and $\lambda(x)$ can be any function. Any two states, picked out by two different choices $\lambda_a(x)$ and $\lambda_b(x)$, are to be identified. This is due to the fact that the probability density $\propto\bar\psi\psi$, which carries the relevant information about the quantum state of the fermion, is not affected by phase shifts $\sim\lambda(x)$, i. e. the probability density $\bar\psi\psi\rightarrow\bar\psi\psi$ is invariant under \eqref{u1-t} (see footnote \ref{fnote}.) 

In the present gauge invariant theory of gravity \eqref{grav-lag} over WIG space, gauge invariance means that the gravitational equations of motion, together with the EOM of the remaining matter fields, are not affected by the gauge transformations \eqref{gauge-t}, which are composed of conformal transformations of the metric $g_{\mu\nu}\rightarrow\Omega^2g_{\mu\nu}$, together with simultaneous gauge transformation of the geometric scalar $\phi\rightarrow\Omega^{-1}\phi$, and appropriate transformations of the remaining fields. Contrary to $U(1)$ gauge symmetry, in this case the fermion spinor $\psi$ and its Dirac's adjoint $\bar\psi$, both share the same conformal weight: $w(\psi)=w(\bar\psi)=-3/2$, so that, under the conformal transformation of the metric, the probability density $\rho_\psi\propto\bar\psi\psi$, transforms like $\rho_\psi\rightarrow\Omega^{-3}\rho_\psi$. The conformal transformation of the metric not only links two different probability densities, but it also links two different metrics, i. e., two different ways of measuring physical time-like $d\tau=\sqrt{-g_{00}}dt$ and space-like $dl=\sqrt{g_{ik}dx^idx^k}$ distances in spacetime. Each one of the conformally related metrics leads to different curvature properties encoded in the curvature tensors: Riemann-Christoffel curvature tensor and its contractions. Hence, choosing a gauge has phenomenological consequences.

Let us, for illustrative purposes, bring to our attention a well-known example which allows to trace a parallel between a spacetime symmetry and gauge invariance. Although the symmetry in the example has nothing to do with gauge invariance, its analysis can give us a sense of overall understanding of our present approach to gauge freedom. Let us briefly discuss about Lorentz invariance of the physical laws, in particular of the EM laws. The laws of electromagnetism, for instance, the inhomogeneous Maxwell equation: $\der_\lambda F^{\mu\lambda}=4\pi j^\mu$, where $F_{\mu\nu}$ is the EM field strength and $j^\mu$ is the current density, are invariant under the Lorentz transformations:

\bea &&F^{\mu\nu}\rightarrow\Lambda^\mu_{\;\;\sigma}\Lambda^\nu_{\;\;\lambda}F^{\sigma\lambda},\;j^\mu\rightarrow\Lambda^\mu_{\;\;\sigma}j^\sigma,\nonumber\\
&&dx^\mu\rightarrow\Lambda^\mu_{\;\;\nu}dx^\nu,\;\eta_{\mu\nu}\rightarrow\Lambda^\sigma_{\;\;\mu}\Lambda^\lambda_{\;\;\nu}\eta_{\sigma\lambda},\label{lorentz-t}\eea where $\eta_{\mu\nu}$ represents the Minkowski metric and $\Lambda^\mu_{\;\;\nu}$ is the Lorentz boost. The latter links two different inertial reference frames (IRFs). Invariance of the EM laws under \eqref{lorentz-t} means that these laws are the same in every reference frame. Yet, the observers in the different IRFs notice different descriptions of the same physical phenomenon. It may happen, for instance, that in a given IRF -- say the rest frame -- the observer, which is equipped with appropriated measuring instruments, measures the electric field with components $E_i=F_{i0}$, exclusively, since the remaining space-space components of the field strength vanish: $F^{ij}=0$. Under \eqref{lorentz-t} the following specific transformations take place: $\tilde F^{ij}=\Lambda^i_{\;\;0}\Lambda^j_{\;\;k}F^{0k}$. Hence, provided that $\Lambda^i_{\;\;0}\neq 0$ and $\Lambda^j_{\;\;k}\neq 0$ for some $i,j,k$, the space-space components of the field strength in the inertial reference frame marked with the tilde do not vanish: $\tilde F^{ij}\neq 0$. In consequence, an observer in the IRF distinguished by the tilde, would measure not only the electric field, but also the magnetic field with (at least one) nonvanishing component $\tilde B_i=\epsilon_{ijk}\tilde F^{jk}/2\neq 0$, where $\epsilon_{ijk}$ is the Levi-Civita symbol. The answer to the following question: Does the magnetic field really exist? depends on the observer. The observer in the rest IRF does not measure any magnetic field so that, to this observer the magnetic field does not exist. However, for the observer in the IRF which we marked with the tilde, the magnetic field exists and it may be used to trigger other physical phenomena. 

We can trace a parallel between the discussed situation, which is derived from Lorentz invariance, and gauge invariance of the gravitational laws: A similar question arises in our analysis of three different gauges in Section \ref{sect-wormholes}: Does the Schwarzschild black hole actually exist? or are its conformal solutions: the wormhole or the naked singularity, the ones that actually exist? As a matter of fact the question is not well posed. Both the black hole, the wormhole and the naked singularity exist in their respective gauges. Hence, the observers ``living'' in the GR gauge see the Schwarzschild black hole, while the observers living in the second gauge see a wormhole and the observers in the third gauge see a naked singularity. The meaningful question is: Which one of these gauges is the one which better reproduces the existing amount of observational evidence? The answer can be given by the experiment exclusively. 

In general we can establish a similarity relationship between the pairs ${\cal I}_\text{irf}=$(IRFs, Lorentz transformations) and ${\cal I}_\text{gauge}=$(gauges, gauge transformations). Let us list several overall similarities.

\begin{itemize}

\item While Lorentz invariance leads to the EM laws being the same in any IRF, gauge invariance leads the gravitational laws to be the same in any gauge. 

\item Above we have given an example of the known fact that the physical description and measured quantities are different in different IRFs. In a similar way, in the present paper we have shown that, the physical description and measured quantities are different in different gauges. 

\item The rest frame is to ${\cal I}_\text{irf}$ what the GR gauge is to ${\cal I}_\text{gauge}$. 

\end{itemize}

There are clear differences between ${\cal I}_\text{irf}$ and ${\cal I}_\text{gauge}$, as well.

\begin{itemize}

\item Lorentz invariance is a spacetime symmetry while gauge invariance is not a spacetime symmetry, since gauge transformations do not modify neither the spacetime events nor the spacetime coordinates. These act only on the fields.

\item Each IRF is associated with one observer (two different observers are linked with two different IRF-s,) but all of the IRF-s live in a same ``Universe.'' Meanwhile, each gauge is populated not only by all possible IRF-s plus all possible general (non inertial) reference frames, but also by all of the SM particles and fields in our world. In a sense we can set an isomorphism between different gauges and different ``Universes.''

\end{itemize} 

Let us further develop the idea stated in the last item above. In a sense each gauge represents a copy of our Universe. Each copy is driven by the same gravitational laws but the way measurements are performed differs from copy to copy. We can imagine the conformal equivalence class ${\cal K}$ -- see the definition in equation \eqref{ce-class} -- as consisting of all $N$ copies of our Universe ($N\rightarrow\infty$). The copies are submitted to the same gravitational laws but to different physical and geometrical descriptions. This interpretation of gauge freedom was called in \cite{quiros-prd-2023-1, quiros-arxiv} as the ``many-worlds'' approach. The main thesis is that the Universe we and the rest of the SM particles and fields live in, may be identified with one of these gauges or worlds. The role of the experiment is to determine, precisely, which one in the infinity of gauges is the one where we live in.


\subsection{The role of the gauge invariants}


Which is the role of gauge invariant quantities in our approach to gauge symmetry? In the first place, notice that the Einstein's tensor of WIG space: $\hat G_{\mu\nu}$, is already a gauge invariant quantity. This leads to gauge invariance of the gravitational equations: $\hat G_{\mu\nu}=6T^\text{mat}_{\mu\nu}/\phi^2$, where $T^\text{mat}_{\mu\nu}$ is the stress-energy tensor of background matter. Let us take, as an illustration, the Friedmann-Robertson-Walker (FRW) line element with flat spatial sections: $ds^2=-dt^2+a^2(t)\delta_{ik}dx^idx^k$, where $t$ is the cosmic time and $a(t)$ is the dimensionless, time-dependent scale factor. In this case, the above EOM plus the continuity equation amount to the following independent differential equations \cite{quiros-prd-2023-1}:

\bea &&3\left(\frac{a'}{a^2}+\frac{\phi'}{a\phi}\right)^2=\frac{1}{\phi^2}\rho_r,\label{frw-eom}\\
&&\rho'_r+4\frac{a'}{a}\rho_r=0,\label{frw-cons-eq}\eea where the tilde means derivative with respect to the conformal time $\tau=\int dt/a(t)$ and we have considered radiation with energy density $\rho_r$ as the background matter. The advantage of considering the conformal time $\tau$ instead of $t$ relies in the fact that the former -- being a coordinate time -- is not modified by the gauge transformations \eqref{gauge-t}. Straightforward integration of \eqref{frw-cons-eq} yields: $\rho_r=\mu_0^2/a^4$, where $\mu_0$ is an integration constant which is not transformed by the gauge transformations \eqref{gauge-t}. Then, if introduce the gauge invariant variable $v\equiv\phi a$, equation \eqref{frw-eom} can be written in the following manifest gauge invariant way: $v'=\mu^2_0/\sqrt{3}$. Integration of this equation leads to the following gauge invariant expression:

\bea v(\tau)=\phi(\tau)a(\tau)=\frac{\mu^2_0}{\sqrt{3}}\tau,\label{frw-eom'}\eea where, without loss of generality, we have considered a vanishing integration constant. There are not other independent differential equations in the gravitational EOM, so that most we can determine, after appropriate mathematical handling of the equations of motion, is the gauge invariant combination $v=\phi a$. Different choices, either of the free function $\phi(\tau)$ or of the scale factor $a(\tau)$, define different gauges. Let us assume, for instance, that $a_\text{gr}(\tau)$ is a known solution of the cosmological GR EOM. Then, since $v$ is a gauge invariant, we have that: $v(\tau)=\phi(\tau)a(\tau)=a_\text{gr}(\tau)$, where we assumed that $\phi_\text{gr}=1$. From this equation it follows that $a(\tau)=\phi^{-1}(\tau)a_\text{gr}(\tau)$. Hence, different choices of the geometric scalar $\phi(\tau)$ generate different behaviors of the scale factor which are conformal to that of the GR known one.

The gauge invariant quantities are also useful in other contexts. Take, for instance, the Kretschmann invariant: $K=\hat R^{\sigma\mu\lambda\nu}\hat R_{\sigma\mu\lambda\nu}$, where $\hat R^\sigma_{\;\;\mu\lambda\nu}$ is the curvature tensor of $\tilde W^\text{int}_4$. The quantity $\tilde K=\phi^{-4}K$, is not only invariant under general coordinate transformations, but it is also invariant under the gauge transformations \eqref{gauge-t}. Hence, if we know, for instance, the static, spherically symmetric GR Kretschmann scalar $K_\text{gr}=48m^2/r^6$, we have that (we chose $\phi_\text{gr}=1$): $\tilde K=\phi^{-4}(r)K(r)=K_\text{gr}=48m^2/r^6$, which leads to

\bea K(r)=\frac{48m^2\phi^4(r)}{r^6}.\nonumber\eea Hence, different choices of the free function $\phi(r)$, define different gauges with given expressions for the Kretschmann scalar. In Section \ref{sect-wormholes} we have used this procedure involving the gauge invariants, in order to get the expressions of the Kretschmann scalar.


\section{Conclusion}\label{sect-conclu}


In the present paper we have investigated the gauge invariant theory of gravity given by the gravitational Lagrangian ${\cal L}_\text{grav}$ in \eqref{grav-lag}, which is based in WIG space. The derived EOM is given by equation \eqref{vac-eq}. With the help of a generalization of the Birkhoff's theorem, we focused in finding static, spherically symmetric solutions to the latter EOM. The resulting scenario allowed us to discuss about the many-worlds approach to gauge freedom -- put forth in \cite{quiros-prd-2023-1} and explored in \cite{quiros-arxiv} within the cosmological framework -- from the local perspective. We confirmed that local experiments: Earth-based and Solar system experiments, specifically the redshift and local position invariance experiments \cite{will-lrr}, are able to rule out several gauges in the related conformal class ${\cal K}$. The hope is that the increasing amount of experimental data, coming both from local and cosmological scale experiments, will allow us to pick out, within the infinity of gauges in ${\cal K}$, the gauge which better describes our Universe.


{\bf Acknowledgments.} The author acknowledges FORDECYT-PRONACES-CONACYT for support of the present research under grant CF-MG-2558591.  





\end{document}